# IMPLEMENTATION OF AI/DEEP LEARNING DISRUPTION PREDICTOR INTO A PLASMA CONTROL SYSTEM

William Tang, Princeton University, PPPL
Princeton Plasma Physics Laboratory, Princeton, New Jersey, 08543 USA
Email: wtang@pppl.gov

Ge Dong[1], Jayson Barr,[2] Keith Erickson[1], Rory Conlin[1], M. Dan Boyer[1], Julian Kates-Harbeck[1], Kyle Felker[1], Cristina Rea[3], Nikolas C. Logan[4], Alexey Svyatkovskiy[1], Eliot Feibush[1], Joseph Abbatte[1], Mitchell Clement[1], Brian Grierson[1], Raffi Nazikian[1], Zhihong Lin[5], David Eldon[2], Auna Moser[2], Mikhail Maslov[6]

*1)Princeton Plasma Physics Laboratory, Princeton, New Jersey, 08543 USA; 2)General Atomics, San Diego, California 92186, USA; 3)Massachusetts Institute of Technology, PSFC, 77 Massachusetts Ave., Cambridge, MA 02139; 4)Lawrence Livermore National Lab, Livermore, California 7000 East Ave., Livermore, CA 94550; 5)University of California Irvine, Irvine, California 92717, USA; 6) EUROfusion Consortium, JET, Culham Science Centre, Abingdon, OX14 3DB, UK*

**Abstract**

This paper reports on advances to the state-of-the-art deep-learning disruption prediction models based on the Fusion Recurrent Neural Network (FRNN) originally introduced a 2019 NATURE publication. In particular, the predictor now features not only the "disruption score," as an indicator of the probability of an imminent disruption, but also a "sensitivity score" in real-time to indicate the underlying reasons for the imminent disruption. This adds valuable physics-interpretability for the deep-learning model and can provide helpful guidance for control actuators now that it is fully implemented into a modern Plasma Control System (PCS). The advance is a significant step forward in moving from modern deep-learning disruption prediction to real-time control and brings novel AI-enabled capabilities relevant for application to the future burning plasma ITER system. Our analyses use large amounts of data from JET and DIII-D vetted in the earlier NATURE publication. In addition to "when" a shot is predicted to disrupt, this paper addresses reasons "why" by carrying out sensitivity studies. FRNN is accordingly extended to use many more channels of information, including measured DIII-D signals such as (i) the "n1rms" signal that is correlated with the n =1 modes with finite frequency, including neoclassical tearing mode and sawtooth dynamics; (ii) the bolometer data indicative of plasma impurity content; and (iii) "q-min" – the minimum value of the safety factor relevant to the key physics of kink modes. The additional channels and interpretability features expand the ability of the deep learning FRNN software to provide information about disruption subcategories as well as more precise and direct guidance for the actuators in a plasma control system.

1. INTRODUCTION

As background for the present studies, we note that in toroidal plasma devices, disruptions are large-scale plasma instabilities that release the plasma stored energy and diminish the plasma current within a very short time-scale [1]. The large energy and particle flux involved can seriously damage the experimental devices, especially when stored energy increases in high-performance plasma experiments (shots) in modern tokamaks such as DIII-D [2] and JET [3], and future tokamak devices such as ITER [4, 5]. In particular, statistical machine learning based predictors relevant to disruptions have become increasingly prevalent in recent years [6,7]. It is particularly noteworthy that in addressing this long-standing challenge, neural networks, which were considered for decades [8], have recently taken a dramatic step forward with the advent of more powerful artificial intelligence approaches enabled by the rapid advances in high performance computing technology at major supercomputing centers. For example, deep learning models based on the Long-Short Term Memory (LSTM) recurrent neural network (RNN) [9] and temporal convolutional neural networks (TCN) [10] have achieved breakthrough results for cross-machine predictions with the aid of leadership-class high-performance-computing (HPC) facilities [11].

The state-of-the-art deep-learning disruption prediction models based on the Fusion Recurrent Neural Network (FRNN) [9] have been further improved. Here we report the new capability of the software to output not only the "disruption score," as an indicator of the probability of an imminent disruption, but also a "sensitivity score" to indicate the underlying reasons/physics subcategories [12] for the imminent disruption. As an indicator of possible causes for future disruptions, the "sensitivity score" can contribute valuable physics-based interpretability for the deep-learning model results, and more importantly, provide targeted guidance for the actuators when implemented into any modern Plasma Control System (PCS). This achievement represents a significant step forward since the 2018 IAEA meeting in moving from modern deep-learning disruption prediction to real-time control that brings novel AI-enabled capabilities needed for the future burning plasma ITER system [5]. Key findings in this paper help address the basic issue/perception that advanced Machine Learning/Deep learning methods are generally hard to interpret. Results presented here are of course supportable by actual data from JET and DIII-D with much of such data having been previously published/vetted in Ref. [9]. These statistical sensitivity studies help address and answer in addition to "when" a shot is going to disrupt, some compelling reasons associated with disruption physics subcategories to help explain "why" it disrupts.

A new scheme is introduced in which real-time control of actuators can be advanced by AI-enabled disruption predictors. Since these deep learning capabilities were developed by using modern programming languages (i.e., Python) to implement the "Keras" algorithmic scheme (explained in [9]), it has become additionally necessary to develop a "Keras2c" converter [13] to enable integration of the AI-based predictor into a plasma control system (e.g., for DIII-D) which is written in the much older C-language. Associated details are explained later in Section 2 of this paper. It is important to keep in mind that Ref. [9] introduced the first adaptable predictive DL software trained on leadership class supercomputing systems to deliver accurate predictions for disruptions across different tokamak devices (DIII-D in the US and JET in the UK). It featured the unique statistical capability to carry out efficient "transfer learning" via training on a large database from one experiment (i.e., DIII-D) and be able to accurately predict disruption onset on an unseen device (i.e., JET). In more recent advances, the FRNN inference engine has been deployed in a real-time plasma control system on the DIII-D tokamak facility in San Diego, CA. This opens up exciting avenues for moving from passive disruption prediction to active real-time control with subsequent optimization for reactor scenarios.

The workflow for the FRNN software can be readily extended to explore the use of many more channels of information. For example, DIII-D signals that are known to be relevant physics-based channels include: (i) "n1rms" – a signal correlated with n =1 modes with finite frequency (where n is the toroidal mode number), including the neoclassical tearing modes (NTM's) and sawteeth dynamics; (ii) bolometer data reflecting the impurity content of the plasma; and (iii) "q-min" – the minimum value of the safety factor directly relevant to important physics such as the kink modes. These considerations motivated including the associated channels directly into the deep learning workflow with the goal of clearer identification of the physics most responsible for the dangerous disruption events with associated guidance for the control actuators. The potential for significant improvement over existing traditional algorithms targeting these signals for plasma condition and disruption control comes from the fact that our AI/deep-learning models are set up for carrying out supercomputing-enabled hyperparameter tuning enhancements of statistical accuracy for complex physical systems with huge feature size without the necessity of "feature engineering." This enables the capability to deliver predictions for unseen conditions, such as new plasma parameters associated with projected larger devices. Moreover, as an indicator of possible causes for future disruptions, the distribution of the "sensitivity score" can provide valuable physics-based interpretability for the deep-learning model results, and more importantly, contribute targeted guidance for the actuators when implemented into any modern PCS. Progress toward this goal represents a significant step forward in moving from modern deep-learning disruption prediction to real-time control that brings novel AI-enabled capabilities with significant beneficial features for deployment in the future on the burning plasma ITER system. Results indicate, for example, that the core radiation power and the familiar MHD safety factor at the radial location near the plasma periphery ($q$-$95$) can represent sensitive channels contributing to physics reasons associated with disruption prediction for specific cases of interest. Significant variability can of course arise when considering general trends in a large overall database.

This paper demonstrates that the FRNN software can be readily extended to using many more channels of relevant information. The potential for significant improvement over existing traditional algorithms targeting these signals for plasma condition and disruption control comes from the fact that AI/deep-learning models have the distinct advantage of being able to statistically enhance predictive accuracy using HPC-assisted modern hyperparameter tuning with the associated training carried out on path-to-exascale supercomputers at leading facilities such as ORNL, ANL, and LBNL in the US. This enables the capability to deliver predictions for as-yet-unseen conditions that can arise, such as new plasma parameters including new DIII-D experiments and those associated with future larger devices such as ITER.

Overall, the key point made in this paper is that when more physics-related channels are statistically included, key insights can be gained on the mechanisms contributing to disruptions. Accordingly, in addition to providing a "disruption score," the present studies compute a "sensitivity score" for each physics-connected channel (as illustrated in Fig. 2). In addition to studying the physics in subcategories of disruptions, these "sensitivity scores" for each channel can in turn provide guidance to the PCS with more precise and direct information for the actuators. Moreover, another important advantage of DL-enabled predictive capabilities is the ability to carry out forecasts significantly earlier in the evolution of the plasma state under consideration. For example, with the information included and the minimum prediction-time threshold set to 30 ms, an improved average alarm time of 100 ms was achieved for FRNN disruption prediction.

This paper also highlights the first results of the implementation of a FRNN LSTM-based deep-learning model into the DIII-D PCS. The real-time computation performance of the FRNN inference engine during DIII-D start-up runs is shown to be compatible with the PCS requirements, which demonstrates that FRNN deep-learning models are entirely capable of disruption prediction tasks to aid control efforts. We also highlight here some recent off-line FRNN results, including a new training scheme with more physics-based signals to improve FRNN disruption prediction capabilities and a new FRNN software suite to compute real-time "sensitivity-scores" for the interpretation and of FRNN deep-learning model disruption predictions with results, for example in Fig.3, displayed later in this paper. The remainder of the paper is organized as follows: in section 2, we provide details of the implementation of the FRNN deep-learning based model into the DIII-D PCS; in section 3, we discuss new FRNN training and disruption prediction results when, for example, the physics-related "n1rms" signal is included as input; in section 4, we introduce the design and output of the "sensitivity scores", and in section 5, we summarize the recent advances and planned future developments of the FRNN software suite.

**2.0 IMPLEMENTATION OF FRNN DEEP-LEARNING-BASED MODEL INTO DIII-D PCS**

The DIII-D PCS [14] is a comprehensive software/hardware system used for real-time data acquisition and feedback control of numerous actuators on the DIII-D tokamak. It regulates many plasma characteristics including shape, position, divertor function, and core performance with a platform for incorporating new control algorithms that are verified and validated. For example, a disruption predictor using the shallow machine learning "random forest" method has recently been implemented and tested [6,7]. As a new category of such algorithms, the AI/deep learning FRNN software has now been integrated into this PCS and has demonstrated successful operation for a significant number of DIII-D shots in the past year (as illustrated for example in Figs. 2 through 5 in this paper). This implementation consisted of four parts: (i) pre-shot configuration; (ii) real-time data collection; (iii) processing through a Keras2c interpreter (see Section 2 below); and (iv) collection of results with associated documentation/publication. The PCS includes a complete user interface allowing an operator to easily choose configuration parameters specific to an algorithm. In the case of FRNN, the required configuration is comprised of a list of normalizing factors applied to each input (as shown specifically in Ref. [9]). These factors are set before the shot and applied during the real-time data collection phase. This process combines heterogenous data from multiple sources during each real-time cycle that includes diagnostics, sensor measurements, and internal calculations from other algorithms.

The data flow of the FRNN software utilizes the normalized measured temporal 0D (magnitude-only) and 1D (spatial) signals as *inputs*. As explained in Ref. [9], the 1D inputs are processed by a set of convolutional neural nets and then concatenated with the 0D inputs to form the input features for the long short-term memory (LSTM) network as well as for the temporal convolutional neural (TCN) network discussed in Ref. [10]. The *output* from either the LSTM or the TCN is *the disruption score which indicates the proximity of the coming disruption event*.

In general, to account for significant control room adjustments such as possible recalibration of a particular diagnostic, associated modifications can include the application of pre-shot normalizing factors to match the offline functions used when training the model. It is important to note here that *the values submitted during real-time correlate to values trained offline*. Finally, the collection of data inputs from the experiment are inserted into a pre-defined Keras2c input data format for use in the generic Keras processor shared by multiple algorithms. The Keras processor produces a predictive result which is then stored post-shot and published in real-time for any interested consumer. More specifically, "Keras2c" is a Python/C library for converting complex Keras/Tensorflow neural networks such as the AI/deep learning FRNN software to real-time compatible C code. Prior to the experiment, a python script parses the trained neural network to extract the necessary parameters and determine the connectivity between layers and nodes. It generates a custom C function to duplicate the forward pass through the neural network. The generated C code makes use of a small C-backend that re-implements the core functionality of Keras/Tensorflow in a safe manner that allows ease of deployment into complex control systems such as the DIII-D

PCS.  "Keras2c" also automatically tests and verifies the correctness of the generated code. *The conversion and testing process of "Keras2c" is fully automated, providing a significant advantage over previous attempts to use neural networks within demanding control applications*.  This enables avoiding the need to either use large non-deterministic software libraries or to code the entire network by hand, and thereby avoiding generating code that proves to be difficult to verify and maintain [13].

At this point, it is useful to highlight the current progress of the FRNN software that has successfully been run on the DIII-D PCS, including establishing a specific category to avoid potential conflicts with other algorithms.  It has now demonstrated reproducible timing for representative real-time shots in the 1ms time-range.  Moreover, the "Keras2c" infrastructure has now been further upgraded to enable sharing the implementation across multiple algorithms (i.e., as many as four) to provide flexibility to address/remediate a number of integration issues that can arise within the PCS.  Finally, it is significant to note that in order to keep pace with attractive emerging technological advances, this AI/DL project has recently leveraged engagement with industry (NVIDIA and Concurrent) to build a unique system that enhances conventional CPU's (central processing units) computer operations with modern GPU's (graphics processing units).  For example, integration of the most advanced NVIDIA A100 GPU is now completing the hardware testing phase.  In future application studies, this can be expected to enable quicker and more efficient examination of the benefits of deploying our current as well as possible new AI/DL algorithms.

**3.0 PHYSICS-BASED SIGNALS FOR IMPROVING DISRUPTION PREDICTION**

As described in the previous section, the data flow of the FRNN software utilizes the normalized measured temporal 0D (magnitude-only) and 1D (spatial) signals as <u>inputs</u> with the 1D inputs systematically processed by a set of convolutional neural nets and then concatenated with the 0D inputs to form the input features for the long short-term memory (LSTM) network [9] as well as for the temporal convolutional neural (TCN) network [10].  The <u>output</u> from either the LSTM or the TCN is <u>*the disruption score which indicates the proximity of the coming disruption event*</u>.

Including physics-based signals as inputs to our deep learning based models are found to improve predictive capabilities. For example, the finite frequency $n=1$ mode amplitude is an important physical quantity, where n is the toroidal mode number.  We note here that although the "n1rms"— as well as "n2rms" and "n3rms"— signals are post-processed non-causal data, "n1rms" can help demonstrate the usefulness of including such instability-related signals in deep-learning based models by shifting the "n1rms" signal input in time by 20 ms to prevent the model from seeing any future information about the plasma [private communication, S. Munaretto, DIII-D magnetic diagnostic system expert, (2019)].  This can be viewed as a useful example to explore how properties of important plasma instabilities, including the kink-like modes and neoclassical tearing modes – can eventually "lock" to the inner wall of the device and lead to disruptions.  In general, the n1rms is the well-defined DIII-D signal point name that represents the $n=1$ finite frequency magnetic perturbations.  This is likewise the representation respectively for the $n=2$, $n=3$, … finite frequency magnetic perturbation.

When the $n=1$ finite frequency mode amplitude is included as an input channel in FRNN, the disruption prediction results can be improved significantly at both the low false positive rate regime and the high false positive regime, as shown in Fig. 1. We performed hyperparameter tuning, as introduced in [9], for FRNN models that are trained with and without the "n1rms" signal and reported the performance of the models that achieved the highest area under the so-called ROC curve on the validation set respectively.  A receiver operating characteristic curve, or ROC curve, is a graphical plot that illustrates the diagnostic ability of a binary classifier system as its discrimination threshold is varied.  It is created by plotting the <u>true positive rate</u> (TPR) against the <u>false positive rate</u> (FPR) at various threshold settings. The true-positive rate is also known as <u>sensitivity</u> or *probability of detection* in <u>machine learning</u>. The false-positive rate is also known as *probability of false alarm*.

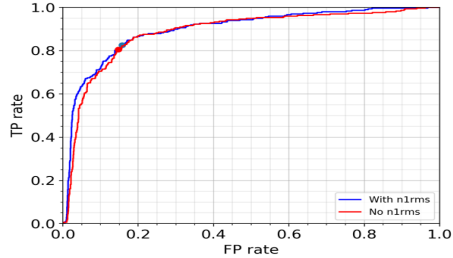

*FIG. 1. Comparison of the ROC curves with and without the n=1 finite frequency mode amplitude ("n1rms"). The vertical (TP/"true positive" rate) or "sensitivity" is plotted vs. the horizontal (FP/"false positive rate) or "probability of false alarm" at various threshold settings.*

More importantly, at the optimal alarm thresholds as indicated by the solid dots in Fig. 1, the model trained with n=1 finite frequency mode signal can raise earlier disruption alarms than the model trained without the n=1 finite frequency signal. Both mean and median of the alarm lead time are increased by more than 100 ms. To demonstrate that the neural network can effectively learn information from the n=1 finite frequency mode signal and thus provide earlier disruption alarm, an example shot from DIII-D (shot #161362) is shown in the left panel in Fig. 2. At around 1.9s, the FRNN model with "n1rms" as an input channel raised the disruption alarm following the onset of the n=1 mode. Before 2s, the "n1rms" signal diminishes while the locked mode amplitude rises up. During this time, the FRNN model trained with the "n1rms" signal provides continuous outputs of disruption alarms. The FRNN model trained without the n=1 mode signal raises a disruption alarm here around 40 ms before the actual disruption, around 200 ms later than the model trained with the n=1 mode signal.

Associated studies of ROC curves using "q-min" and bolometer signals have not been adequately completed at this time due to limited data availability issues. For example, while a large portion of the shots studied did have strong q95 signals (with q95 defined as the safety factor near the plasma edge), they had very little q-min (minimum q) and q-profile information, thereby leading to a much smaller statistical database to analyze. This was also the case for the limited bolometer data which also had associated complexity complications with proper normalizations.

In any case, we note that with the other basic plasma quantities as input, the n=1 finite frequency signal usually does not confuse the model when the mode is not leading to disruptions. For example, in shot #170239, as shown in the right panel of Fig. 2, although a tearing mode appears at 2-3s, the FRNN output remains at a constant low level. These results demonstrate the application of neural networks to enable discriminating between those tearing modes that can be linked to the onset of disruptions (e.g., DIII-D shot number 161362) and those that are not observed to do so (e.g., DIII-D shot number 170239).

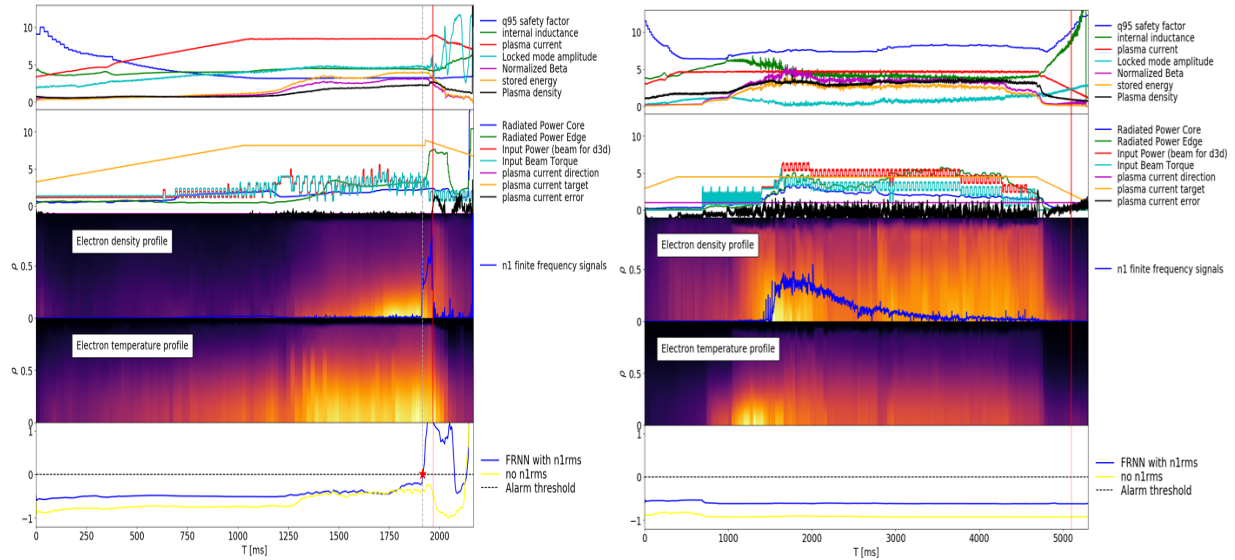

*FIG. 2. DIII-D shot number 161362 in the left panel (example of a disruptive case) and DIII- shot number 170239 (example of a non-disruptive case) in the right panel. In each panel, the upper 4 sub-panels show measured signals as FRNN input, and the bottom sub-panel show FRNN model outputs.*

Overall, it has been well established from many experimental observations in tokamaks (e.g., Ref. [12]), that NTM activity is important disruptions, and this is in fact a key motivation for the studies in this section of our paper. It is important to keep in mind though that there have been many DIII-D shots observed where the NTM islands become large and even "locked" – but with where *disruption is actually avoided*. Tearing modes and locked modes are in fact known to have very complicated dynamical behavior that closely relates to the disruptivity of the plasma. So, an actual specific "threshold" for NTM's is quite difficult to define. We have accordingly introduced in these studies systematic AI/deep learning *statistical* methodology that incorporate n=1 mode activity – instead of a simplistic ill-defined "locked mode threshold" to help forecast disruptions.

### 4.0   REAL-TIME SENSITIVITY STUDIES
To interpret the disruption predictive capabilities of the DL-based model, we have developed sensitivity study schemes for individual test shots -- schema that can be implemented in real-time along with the regular FRNN model inference engine as introduced in Sec. 2. For each shot, the sensitivity study helps answer the question of "why the neural network outputs a high disruption score and raises a disruption alarm at a given time?" More importantly, the results from the sensitivity study scheme can provide detailed indications of which physical quantities can provide relevant proximity guidance for disruptive scenarios, and this information may directly aid control efforts by identifying the appropriate actuator to possibly optimize "proximity to a disruption." This aspect is further noted in comments in Section 5.0 and are currently being planned in collaboration with J. L. Barr, et al. (Ref. [15]) for future AI/DL FRNN control investigations that include the deployment of actuators.

### 4.1 Calculation of the sensitivity score

In Ref. [9], the authors provided results from studies examining the importance of a signal to illustrate the contribution of each such physical signal to the test results of the entire test database. In these signal-importance studies, the model is re-trained for a "c" number of times with each physical signal excluded from the training and test database, where "c" is the number of physical signals, and the test result is reported in comparison with the baseline where all signals are included. In the present paper, we have included all signals during training. In the course of testing for each shot, we have performed inference "c" times in parallel – such that at each time, one physical signal is suppressed to output a "c" number of disruption scores. The sensitivity score of each signal is defined as the absolute difference between the baseline output (where full input is used) and the output where the signal of interest is suppressed in the test data. For the standard trained models, we suppress a signal by replacing that signal with a fixed typical value from non-disruptive shots. For noise-aware models which are trained with dropped out signals [8], and are 'familiar' with all-zeroes-signals, we suppress a signal by directly replacing it with zeros. The outputs of these two schemes are generally qualitatively consistent, showing the robustness of the sensitivity study results with respect to the replacement values.

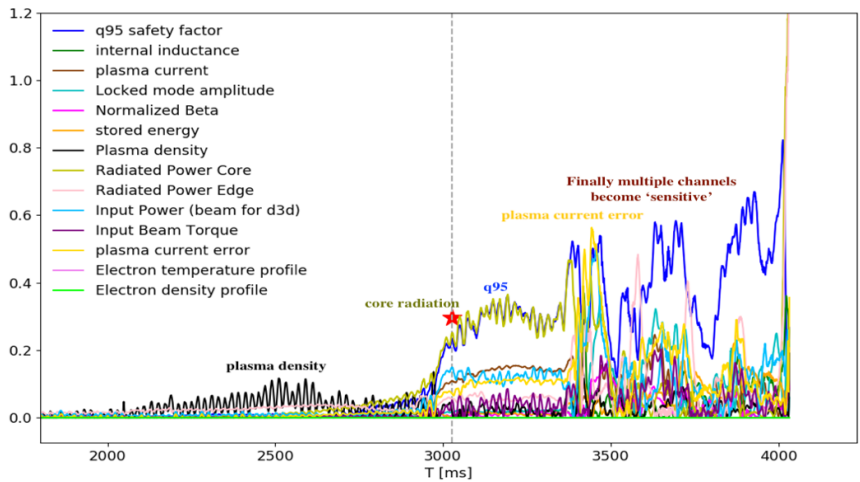

FIG 3. Illustration of the contributions of physics-based signals to the temporal evolution of the sensitivity score (dimensionless) for disruptions that is plotted on the y-axis for the representative DIIID shot #164582.

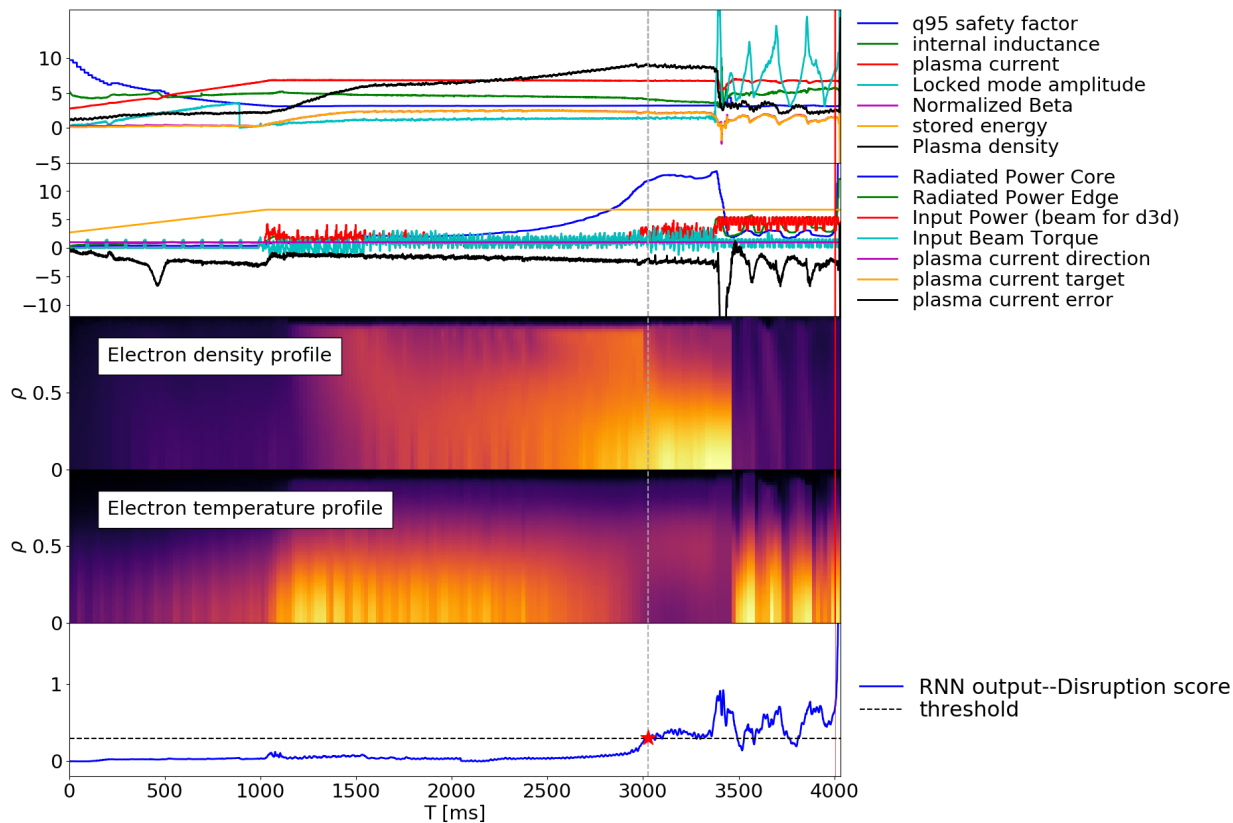

FIG 4. Evolution of each normalized physical signal for DIII-D shot #164582 in the upper 4 panels. The bottom panel shows the time history of the FRNN output

An example of the sensitivity study result of DIII-D shot #164582, Fig. 3 illustrates where the evolutions of the sensitivity scores of each signal are plotted as a function of time. At around 2 to 2.7 seconds, the disruption score slowly rises up, as shown in the last panel in Fig. 4, and the plasma density is shown as the most sensitive channel. Experimentally during this time, the plasma density gradually rises due to impurity influxes, as shown in the first panel as a black line in Fig. 4. The influx of impurities is followed by the rise of the core radiation, which leads to disruption alarm at around 3s.

At the disruption alarm time, the channels with the highest sensitivity scores are core radiation and q-95. Such sensitivity studies could accordingly be used in future studies to suggest PCS operations for example that might engage ECH in the core region to flush impurities or to simply raise q-95. We also note here that at around 3.5s, a large tearing mode starts to develop, possibly due to the high impurity level. This in turn leads to mode locking as shown by the onset of the locked mode amplitude in the first panel of Fig. 3. Around this time, every channel becomes sensitive, and the sensitivity scores begin to change rapidly as the plasma current error rises.

**4.2 Sensitivity score from "Zero-value Replacement Procedure" for "noise-aware" models**

We show the sensitivity study result for DIII-D shot #162975 in Fig.5, using the "zero-value replacement procedure" for individual channels during inference studies using noise-aware models. It is illustrated that after 1.5 seconds, the disruption score begins to increase and raises a disruption alarm. The sensitivity score of different signals at the alarm time is shown in the lower panel of this figure as an indication of their contributions to the disruption alarm. Sensitive channels including internal conductivity, the q95 safety factor, and plasma density all can contribute to a general deterioration of plasma shape and plasma control. This interpretation is qualitatively consistent with the observation of experimental characteristics. For example, at around 1.3 s the plasma shape begins to alter as the X point moves off-target in this shot. Future collaborative studies with EFIT [13] can help to ascertain the magnitude of such changes. In Fig. 5, a tearing mode begins to develop at 1.45s -- which then plausibly evolves into the locked mode at round 1.6-1.7s when the disruption alarm rises. It is important and significant to highlight the fact that in both of the shots that we have analyzed as examples here that *although tearing modes that lock are important causes for the disruptions, the actual locked mode amplitude is not a sensitive signal for both of these shots. Our results indicate that the deep neural network can process the basic plasma information effectively and that the interpretation of the outputs in the form of a real-time sensitivity study can provide early diagnostic information with associated guidance for plasma control and detailed disruption proximity analysis*. This is supported by the fact that we did not use the "n1rms" signals as input for these studies. Accordingly, such sensitivity studies are indeed potentially capable of contributing significantly to real-time disruption mitigation and avoidance investigations.

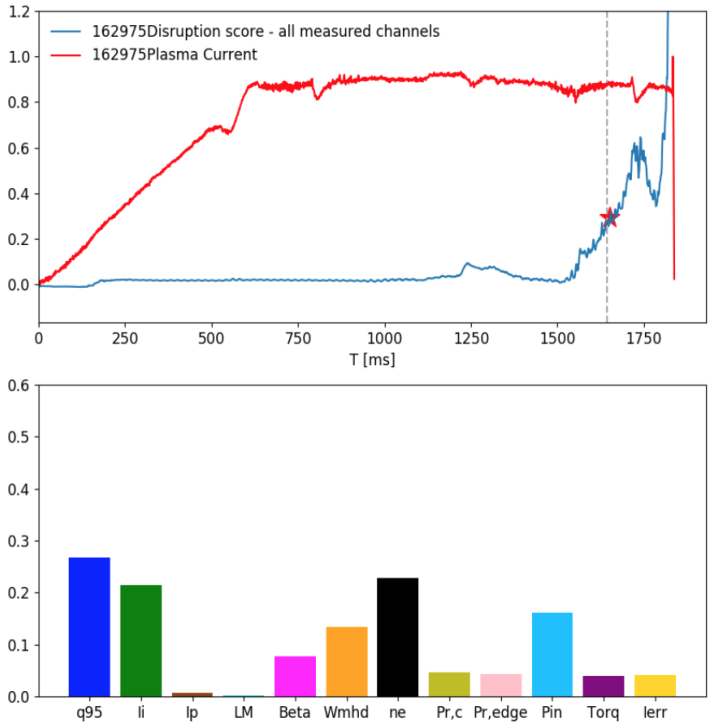

FIG. 5. DIII-D shot #162975. The upper panel shows the evolution of the plasma current (red line) and the FRNN output of the disruption score (blue line -- with the lower panel showing the sensitivity scores (for associated signals such as q95, etc.) at the time of the disruption alarm (red star in the upper panel).

5.0 SUMMARY & ASSOCIATED FUTURE INVESTIGATIONS
In the present paper we have described the implementation of our AI/deep learning disruption prediction software into the DIII-D plasma control system (PCS). This includes the introduction of a new method for interpreting results from deep neural networks using a sensitivity study methodology with significant implications for real-time actionable integration into future plasma control systems.  Key conclusions and consequences from demonstrating the sensitivity of the alarm from the AI/deep learning FRNN disruption predictor include illustrating the interpretability of deep-learning based models and capability to provide physics based information for the PCS when a disruption alarm is raised. The present studies are indicative that if more signals can be included in the training database, it can be expected that better predictive capability in associated future investigations will likely emerge. Here we are highlighting the initial exciting ability of the neural network to discriminate between disruptive and non-disruptive tearing modes.  In the future, additional useful guidance for the PCS could be generated, for example, by carrying out parallel inference studies with small variances for key quantities displayed in Fig. 5 (density, li, …) with results leading to a possible steepest-descent trend in the disruption score that might provide an alternative approach to the current sensitivity measure.  Overall, in planned ongoing and future investigations, we will extend and interconnect stimulating features of this work by providing the deep-learning sensitivity output in real-time into the proximity control architecture designed for handling major disruption causes in the DIII-D PCS [15].

ACKNOWLEDGEMENTS

The authors are grateful to Dr. David Humphreys of General Atomics, DIII-D for his careful review with helpful suggestions for improvement and clarification that have been been integrated into this manuscript. Material for these studies is based upon work supported by the U.S. Department of Energy, Office of Science, Office of Fusion Energy Sciences, using the DIII-D National Fusion Facility, a DOE Office of Science user facility, under Awards DE-FC02-04ER54698; DE-AC02-09CH11466, DE-AC52-07NA27344, DE-SC0020337, DE-SC0014264. This work has been carried out within the framework of the EUROfusion Consortium and has received funding from the Euratom research and training program 2014-2018 and 2019-2020 under grant agreement No 633053. The views and